\documentclass{icrc}

\usepackage{times}
 \usepackage{graphicx} 

\begin{document}

\title{School Cosmic Ray Outreach Detector (SCROD)}
\author[1]{L.A. Anchordoqui}
\affil[1]{Department of Physics, Northeastern University, Boston, MA 02115 USA}
\author[2]{J. Cook}
\affil[2]{Northeastern University and Paige Academy, Boston, MA 02115 USA}
\author[1]{M. Gabour}
\author[1]{N. Kirsch}
\author[1]{J. MacLeod}
\author[1]{T.P. McCauley}
\author[1]{Y. Musienko}
\author[1]{T.C. Paul}
\author[1]{S. Reucroft}
\author[1]{J. D. Swain}
\author[1]{R. Terry}

\correspondence{J. Swain (john.swain@cern.ch)}

\firstpage{1}
\pubyear{2001}


\maketitle

\begin{abstract}
We report on our studies of applying novel detector technologies developed for
LHC-era experiments to cosmic ray detection. In particular, we are investigating usage of 
scintillating tiles with embedded wavelength-shifting fibers and avalanche photodiode
readout as part of a robust, inexpensive cosmic air shower detector. In the near future,
we are planning to deploy detector stations based on this technology at area high schools
and colleges as part of an outreach and education effort, known as SCROD.
\end{abstract}

\section{Introduction}

The project we describe here, known as {\em SCROD} for {\em S}chool {\em C}osmic
{\em R}ay {\em O}utreach {\em D}etector, 
is based on a idea which is simple but has great potential:
install cosmic ray detectors suitable for continuous muon counting and
detection of building-sized (or larger) extensive air showers in high schools and 
relay collected data via the internet to a central repository which
is accessible to all participating schools.  
The principle aim of the project is education, but there is potential for 
contribution to cosmic ray physics as well.  Involving students in a 
project making real measurements in a living field appears to be more likely to 
spark an interest in physics than the usual ritual of 
repeating century old experiments whose conclusions are foregone. 
A number of groups are pursuing similar programs (CROP, CHICOS, ALTA, WALTA)~\footnote{Relevant web pages are:\\ 
{\tt http://www.unl.edu/physics/crop.html} \\
{\tt http://www.chicos.caltech.edu/}\\
{\tt http://csr.phys.ualberta.ca/\~~alta/}\\
{\tt http://www.phys.washington.edu/\~~walta/}\\
{\tt http://www.hep.physics.neu.edu/scrod/}
}
using various approaches and recycling equipment to various degrees.  Here
we discuss our approach and the current status of the work.

\section{Physics Potential}

There are a number of topics which can be addressed with an array of detectors
of the type we are proposing, ranging from searches for long-range correlations
among air showers (perhaps the primary goal) to other topics in astrophysics.

Our pilot phase will take place in Boston, which, with its dense population of
schools, is naturally conducive to a reasonably granular detector.  In 
this phase, therefore, SCROD will function somewhat like earlier experiments
such as AGASA and CASA, sampling the shower front with scintillators.  

Depending on what the highest energy cosmic rays actually are, 
it is conceivable that there exist long-range correlations among
the air showers they produce. 
Either observation or non-observation of such correlations would be
an important result which cannot readily be obtained
except by geographically extensive experiments.

Several processes could give rise to very long-range
correlations. One is 
the Gerasimova-Zatsepin (GZ) effect \citep{GZ,GZwatson,epele}, in which a high
energy atomic nucleus approaches the earth and dissociates on an optical photon from the
sun. The (two or more) nuclear fragments can then reach the earth at distant
locations, but close together in time.  Since the composition of high energy 
cosmic rays is unknown, and its determination from single extensive air showers is
complicated by sensitivities of observables to details of the hadronic interaction
model chosen, it would be interesting to search for such events.

In the GZ effect, the distance by which nuclear fragments are 
separated upon arrival at the earth depends on their deflection in the 
magnetic field of the solar system.
Recent analysis~\citep{GZwatson} for iron nuclei has indicated that very large
separations are to be expected.  For iron nuclei with energy around
1~EeV, for example, most separations wll be in excess of 100~km.  
Clearly, extensive detectors are required to observe such events.
Figure~\ref{fig:cities} shows roughly the expected rate for various
separations (see the appendix for more detail). 
\begin{figure}[htb]
\vspace*{2.0mm} 
\includegraphics[width=8.3cm]{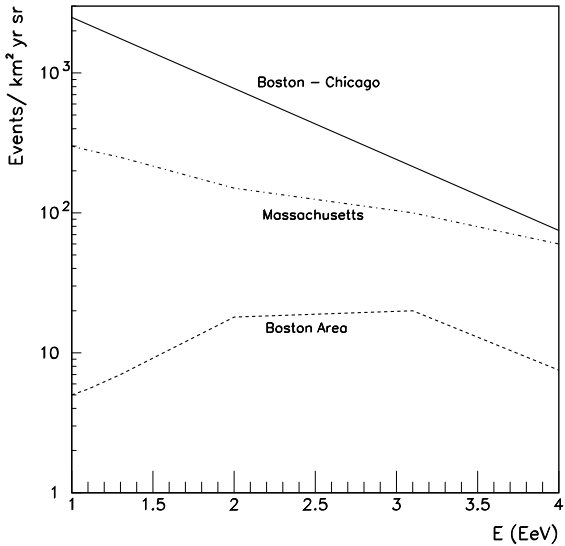} 
\caption{\label{fig:cities}Expected rates versus energy for correlated events 
due to iron nuclei disintegrating on solar photons.  This figure is for the
case of events arriving from a direction close to the sun.  The three
lines show approximate rates characteristic of the separations attainable
in different geographical areas (the Boston area, the entire state of Massachusetts, 
and a separation of roughly the distance from Boston to Chicago.)}
\end{figure}

There are also other conceivable mechanisms to produce a similar effect.
For example highly energetic dust grains could dissociate and give rise
to widely-separated showers~\citep{Anchordoqui}.
One might also conceive of dramatic cosmic events
that may also pepper the globe with many high energy
cosmic rays all at about the same time. With the GPS timing
information, it will be possible
to compare and correlate data taken with SCROD with those taken at neutrino and
gravitational radiation detectors.  Finally we note that 
there is already some suggestion of experimental evidence for long range 
correlations in the literature~\citep{wada,carrel}.

The primary background for genuine long-range correlations will arise from random coincidences
of low energy showers.  This can be controlled by detector spacing, which effectively 
sets an energy threshold, and possibly pulse-height analysis of the scintillator signals.

\section{Education Goals}

The goal is that students, under the advisement of professional physicists and their 
teachers, will be responsible for the day-to-day running of the 
experiment, for the data analysis and search for time correlations, and will in some
cases devise unique projects using their station.  We are also consulting with area teachers
to begin developing ways to use the apparatus to catalyze related classroom activities.
At the current prototype phase, we are involving a few high school students and beginning
undergraduates directly in the development efforts.  Aside from its value to the
students, this helps us to ascertain which
aspects of the project will have to be kneaded into a more pedagogically usable form.

\section{Detector Description}

The hardware proposed for the detector sites consists of the following main components:
1) a set of plastic scintillating tiles with wavelength-shifting fibers; 2) avalanche
photodiodes to read out the fibers; 3) a GPS-based system to time-stamp the signals; 4)
a personal computer (PC) for local data acquisition and 5) the Internet to provide an inexpensive 
wide-area data acquisition system.  A single station will be equipped with 3--5 separate
scintillators, arranged on the school rooftop.  We plan to procure new detector equipment 
(scintillators and fibers) while recycling computers, which otherwise would be the
single most expensive component of the system.  In this way we can deliver a quality
detector at reasonable cost.

\subsection{Scintillating Tiles with APD Readout}

\begin{figure}[htb]
\vspace*{2.0mm} 
\includegraphics[width=8.3cm]{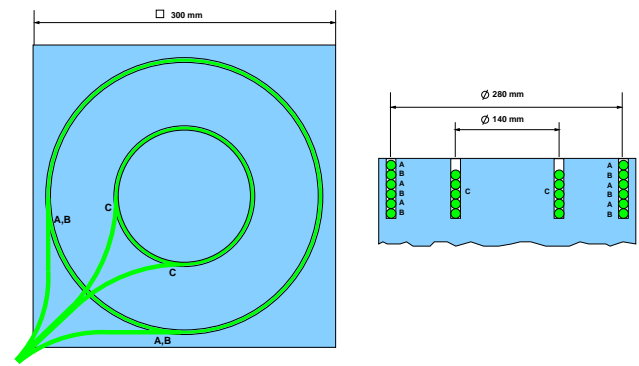} 
\caption{\label{fig:fiberschem}Schematic representations of scintillators with 
embedded wavelength-shifting fibers.}
\end{figure}

The scintillator we use is adapted from technology developed for the 
LHC-b pad/preshower detector~\citep{lhcb}, and comprises a $30 \times 30$~cm
plastic scintillator slab with two circular grooves machined in it.  Three
wavelength--shifting fibers (Bicron BCF-91) are embedded in the grooves, 
1 in the inner grove, 2 in the outer, to shift the scintillation light
into the sensitive region of the readout apparatus and to serve as a 
light guide.  The scheme is illustrated in Figure~\ref{fig:fiberschem}.
The entire assembly is wrapped in white Tyvek~\footnote{Tyvek
is a trademark of DuPont.} paper to increase light collection efficiency.

To read out the fibers, we use an avalanche photodiode (APD).   APDs are essentially
photodiodes with an internal gain mechanism. They can have high quantum efficiencies, exceeding those
of photomultiplier tubes. They are also mechanically robust~\citep{Baccaro,Yuri,Joan} and easy to use, requiring 
a supply of only a few hundred volts, a current-limiting resistor, and a preamplifier.
Furthermore, they require only a few hundred nanoamperes of current to function; an
apparatus which presents no risk of electric shock is attractive for deployment
at schools.  

\begin{figure}[thb]
\vspace*{2.0mm} 
\includegraphics[width=8.3cm]{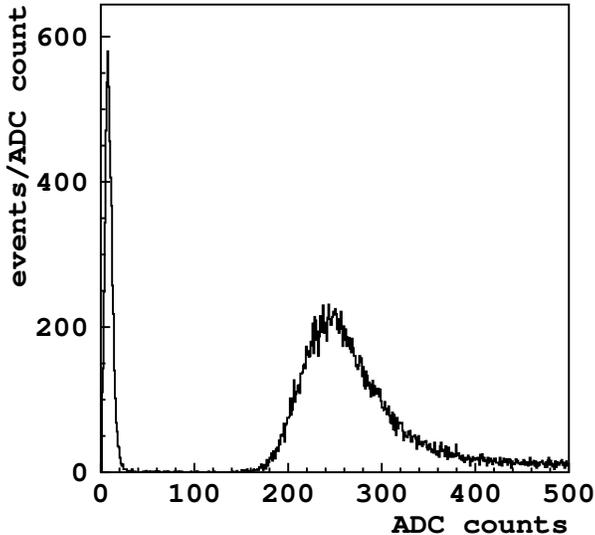}
\caption{\label{fig:pulseheights} Signal due to single muons passing through
the scintillator using APD readout.} 
\end{figure}

Using a low-noise amplifier we have designed together with the 
scintillator-fiber-APD assembly yields a very clean signal, as shown
in Figure~\ref{fig:pulseheights}.

\subsection{GPS Timestamp and Data Acquisition}

\begin{figure}[thb]
\vspace*{2.0mm} 
\includegraphics[width=8.3cm]{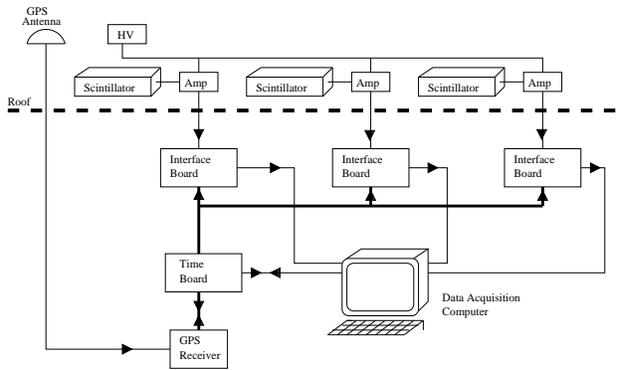}
\caption{\label{fig:layout} Readout scheme for a SCROD station.  The Time Board
broadcasts the time count, using the 1PPS signal from the GPS receiver to 
stay synchronized to universal time.  The Interface Boards latch the time upon
receipt of a signal from a scintillator, then relay it to the computer.} 
\end{figure}

In the current prototype version, the amplifier signal is passed through a 
discriminator to generate a TTL pulse, which is used to latch the time of
each hit.  The time is broadcast from a central board comprised of two
sets of counters, one of which records the number of pulses delivered by
the GPS receiver's 1 pulse per second (1PPS) line, while the other 
is clocked by an on-board 100~MHz oscillator and reset each second 
by the 1PPS line. The timing resolution offered by the fast rising 
edge of the 1PPS signal is about 40~ns.  
  Each scintillator has its own time latching
circuitry which converts the time to serial RS-232 signals and relays
it to a serial port in the computer; there is one serial port for each
scintillator.  This design is reasonably modular so that a given site can
easily attach another scintillator, or in principle any other piece of 
hardware which records data that should be timestamped.
The basic scheme is illustrated in Figure~\ref{fig:layout}.

This design also allows for very simple electronics.  Our prototypes were
constructed by first-year undergraduate students using quite 
inexpensive off-the-shelf components.

To the extent possible, we are devising a purely offline software trigger; all
hits are stored in their respective serial ports and accessed by reader 
programs running as separate threads on the PC.  A triggering module can access the 
data from the different threads and apply trigger logic.
This allows students to develop their own triggering schemes
in a very easy manner, and it also will facilitate future upgrades and uniformity
among all the stations.  Changing the trigger will only entail uploading a new piece of
code, instead of altering the hardware at numerous sites.

\section{Summary}

We are adapting detector technologies from our work on LHC experiments to develop
inexpensive cosmic air shower detectors suitable for deployment at high schools.
The primary goal is to interest students in physics by involving them directly
in a project which has the potential to make some meaningful measurements.  
The essential ingredients have been designed and tested (partly by students)
and construction of a station for deployment is underway.

\appendix
\section{}

The rate of occurrence of these kind of events is given by 
the cosmic ray flux, the 
fragmentation probability, and
the fraction of GZ events with separation 
distance $d < d_{\rm max}$, $f_{d_{\rm max}}$.
The rate of events above a given energy $E$ for a surface detector of 
area $S$ and solid angle $\Omega$ reads,
\begin{equation}
{\rm Rate} (E_{\rm CR} > E) \sim\, \Phi (E_{\rm CR} > E) \,\eta_{\rm GZ}(E)\, 
f_{d_{\rm max}} \,S\,\Omega,
\end{equation}
where $\eta^{\rm Fe}_{\rm GZ}(E) \sim 10^{-5} - 10^{-4}$ is the fragmentation 
probability, and
\begin{equation}
\Phi (E_{\rm CR} > E)\,\sim 47 \left(\frac{{\rm EeV}}{E}\right)^2 
\,\,{\rm km}^{-2}\, {\rm yr}^{-1}\, {\rm sr}^{-1},
\end{equation}
is the measured cosmic ray flux above the knee 
($E_{\rm CR} > 3 \times 10^{15}$ eV).

\begin{acknowledgements}
This work was supported in part by the National Science Foundation and CONICET.
\end{acknowledgements}

\end{document}